\documentstyle[11pt]{article}
\newcommand{\be}{\begin{equation}}
\newcommand{\ee}{\end{equation}}
\newcommand{\ba}{\begin{eqnarray}}
\newcommand{\ea}{\end{eqnarray}}

\begin{document}
\begin{titlepage}
\begin{flushright}
HD--THEP--98--43\\
\end{flushright}
\quad\\
\vspace{1.8cm}
\begin{center}
{\bf\LARGE Hints from the Standard Model}\\
\bigskip
{\bf\LARGE for Particle Masses and Mixings\footnote{Invited
talk given at the meeting on ``What comes beyond the Standard
Model?'', Bled (Slowenia), July 1998}}\\
\vspace{1cm}
Berthold Stech\\
\bigskip
Institut  f\"ur Theoretische Physik\\
Universit\"at Heidelberg\\
Philosophenweg 16, D-69120 Heidelberg\\
\vspace{3cm}
{\bf Abstract:}\\
\parbox[t]{\textwidth}{The standard model taken with a momentum
space cut-off may be viewed as an effective low energy
theory. The structure of it and its known parameters can give
us hints for relations between these parameters. In the present investigation the Higgs
problem will be discussed, the possible
connection of the Higgs meson with the heavy top quark, and the
geometric structure of the quark and lepton mass matrices.}
\end{center}\end{titlepage}
\newpage
\section{Introduction}
The standard model is likely to describe the effective
interaction at low energy of an underlying more fundamental theory.
One may speculate that some parameters which emerge at long
distances are insensitive to details of what is going on at much
higher scales. For instance, their values may be given by the fix
points of renormalization group equations, thus being rather
independent of the starting numbers at small distances \cite{1}. Or, these
parameters could arise in a bootstrap-type scenario \cite{2}.

In this talk I do not want to discuss specific models of this
type, but I will simply look at the measured parameters of the
standard model and at its divergence structure, in order to
find hints for possible connections between these parameters.
I will concentrate on the vacuum expectation value of the Higgs 
field, which is defined without reference to external particles 
and their momenta. Thus its divergence property, depending 
solely on the structure of the vacuum, can be quite 
different from those of ordinary coupling constants, which 
can be renormalized using a momentum subtraction scheme.

By taking the standard model as an effective theory, one should use
a momentum cut-off. The dependence of measurable quantities on
the cut-off reflects the influence of new physics on the low
energy domain. The minimization of this influence provides
suggestions for the sought relations.

\section{The vacuum expectation value of the Higgs field and
the invariant Higgs potential}

We write the Higgs part of the Lagrangian in the form
\ba\label{2.1}
{\cal L}_H&=&(D_\mu\Phi)^\dagger D_\mu\Phi+\frac{J}{2}
\Phi^\dagger\Phi-\frac{\lambda}
{2}(\Phi^\dagger\Phi)^2\nonumber\\
&&+\frac{1}{\sqrt2}j\cdot\Phi\quad+\quad{\rm Yukawa\ couplings}\nonumber\\
\Phi&=&\frac{1}{\sqrt2}\left(\begin{array}{c}
\varphi_1+i\varphi_2\\
\varphi_0+i\varphi_3\end{array}\right).\ea
The quantity $J$ is taken to be $J=J_0+J_1$
with $J_0>0$ describing the Higgs mass parameter responsible
for spontaneous symmetry breaking. $J_1$ can be viewed as an
outside field. It is used for generating a gauge invariant potential 
and will finally be set to zero. The quantity $j$ determines
the field direction of the spontaneous symmetry breaking. It could be
due to a light quark condensate and may be neglected after
the occurrence of the symmetry breaking. Accordingly, the potential
in the tree-graph approximation takes the form
\be\label{2.2}
V_0=\frac{\lambda}{8}(\sum_i\varphi^2_i)^2-\frac{J}{4}
\sum_i\varphi^2_i-\frac{1}{2}j\varphi_0 ~~.\ee
For $J>0$ the minimum of $V_0$ occurs
for $\varphi_i=\hat\varphi_i(J,j)$
with
\be\label{2.3}
\hat\varphi_{1,2,3}=0,\quad \lambda\hat\varphi_0^3-J\hat\varphi_0=j ~~.\ee
We have to select the real solution of the cubic equation. In the
limit $j\to0$, $\lambda\not=0$ one gets
\ba\label{2.4}
\hat\varphi_0(J)&=&\sqrt{\frac{J}{\lambda}},\qquad m^2_H(J)=\frac
{\partial^2V_0}{\partial\varphi^2_0}\big|_{\varphi=\hat\varphi}
=J\nonumber\\
&& m^2_i=\frac{\partial^2V_0}{\partial\varphi^2_i}\big|
_{\varphi=\hat\varphi}=0\qquad~ i=1,2,3~~.\ea
By replacing as usual $\varphi_0(x)$ by
\be\label{2.5}
\varphi_0(x)=\hat\varphi_0(J,j)+H(x)\ee
the interaction part ${\cal L}_{int}'$ of the
shifted Higgs Lagrangian allows one to evaluate $<H>$. To
lowest-order the expression is
\be\label{2.6}
<H>=i\int d^4x<0|T(H(0),{\cal L}_{int}'(x))|0>.\ee
The result \cite{3} obtained from (\ref{2.6}) can be used to write
the gauge-invariant vacuum expectation value of the square
of the Higgs field $\sigma(J)=<\sum_i\varphi_i^2>$ for $J>0$
and $j=0$
in the form
\ba\label{2.7}
\sigma(J)&=&\frac{J}{\lambda}-2<H^2>+2\frac{g^2+{g'}^2}{4\lambda}
<Z_\mu Z^\mu>\nonumber\\
&&+4\frac{g^2}{4\lambda}<W^+_\mu W^{-\mu}>-\frac{g^2_t}{\lambda}
<\bar tt>/m_t ~~.
\ea
Here $g,g'$ are the gauge couplings for the vector bosons
$W$ and $Z$, and $g_t$ denotes the Yukawa coupling for the top
quark. Fermions of lower mass are neglected, but could easily be
added. The ``vacuum leaks'' $<H^2>, <Z_\mu Z^\mu>,...$
could be finite in the full theory with a correspondingly
modified vacuum structure. By taking for the effective theory
a particle momentum cut-off chosen to
be universal for all propagators, Eq. (\ref{2.7}) leads to
\ba\label{2.8}
&&\sigma(J)=\frac{J}{\lambda}+\frac{\Lambda^2}{8\pi^2}2\nonumber\\
&&-\frac{\Lambda^2}{8\pi^2}[3+3\frac{g^2+g^{'2}}{4\lambda}
+6\frac{g^2}{4\lambda}-12\frac{g^2_t}{2\lambda}]\nonumber\\
&&+\frac{J}{8\pi^2}
\left\{1+3\left(\frac{g^2+g^{'2}}{4\lambda}\right)^2
+6\left(\frac{g^2}{4\lambda}\right)^2-12\left(\frac{g^2_t}
{2\lambda}\right)^2\right\}\ln\frac{\Lambda^2}{J/\lambda}
\nonumber\\
&&-\frac{J}{8\pi^2}\left(\ln\lambda+3\left(\frac{g^2+g^{'2}}
{4\lambda}\right)^2\ln\frac{g^2+g^{'2}}{4}\right.\nonumber\\
&&\left.+6\left(\frac{g^2}{4\lambda}\right)^2\ln\frac{g^2}{4}-12\left(
\frac{g^2_t}{2\lambda}\right)^2\ln\frac{g^2_t}{2}\right) ~~.
\ea
The term $\frac{\Lambda^2}{8\pi^2}2$ which represents the free field
part of $<\Sigma_i\varphi^2_i>$ is written separately. 
It also appears in the case $J<0$ where one has $\hat\varphi_0(J)=0$
and no spontaneous symmetry breaking. For $J < 0$ one finds to one 
loop order
\ba\label{2.9}
\sigma(J)&=&\frac{\Lambda^2}{8\pi^2}2+\frac{J}{8\pi^2}
\ln\frac{2\Lambda^2}{-J} ~~.
\ea
Thus, in our approximation\footnote{At $J=0$ the one-loop
approximation for $\sigma(J)$ is 
insufficient. However, (\ref{2.10}) is expected to hold
for the change of $\sigma$ within a larger region around
$J=0$.}, $\sigma(J)$ is discontinuous
at $J=0$, indicating a first-order phase transition with strength
proportional to $\Lambda^2$:
\be\label{2.10}
\Delta\sigma=-\frac{\Lambda^2}{8\pi^2}[3+3\frac{g^2+g^{'2}}
{4\lambda}+6\frac{g^2}{4\lambda}-12\frac{g^2_t}{2\lambda}]\ee
If we would keep $j\not=0$, the jump of $\sigma(J)$
would be replaced by a rapid, but now continuous, change
in the region $-j^{2/3}\leq J\leq j^{2/3}$. The right-hand
side of (\ref{2.10}) with its quadratic divergence also shows up
in the conventional (non-gauge invariant) Higgs potential $V(<\varphi>)$.
Its appearance constitutes an essential part of the hierarchy
problem and necessitates large fine-tuned subtractions  
depending in a specific way on the standard model parameters. 
This is quite unnatural 
and is known as the hierarchy problem. Quadratic divergencies do not 
occur in supersymmetric models where the super partners of each 
particle provide for a cancellation of such divergencies. 
In spite of this elegant solution, 
attempts have been made to obtain a cancellation of divergencies 
by special choices of the standard model parameters which could 
make the introduction of superpartners at the weak scale 
unnecessary \cite{4,3}. It is speculated that the particle couplings 
arrange themselfs such as to stabilize the particle masses. Expressed 
in terms of masses, the relation for which (\ref{2.10})
vanishes is
\be\label{2.11}
S_{\Lambda^2}=(3m^2_H+3m^2_Z+6m^2_W-12m^2_t)/<\varphi_0>^2 = 0 ~~.\ee
It is known as the Veltman condition \cite{4}. A sofar unsolved problem 
is the extension of this relation which includes higher loop effects. 
To my knowledge no gauge invariant scheme is known which allows one to 
isolate the quadratic divergencies with the help of a single cut-off 
parameter (except lattice calculations).

Eq. (\ref{2.8}) can also be obtained in a more general context.
By defining the ``free energy'' $W(J,j)$ as the logarithm of
the partition function with the action given by (\ref{2.1}),
one can obtain $\sigma $ from
\be\label{2.12}
\sigma =-4\frac{\partial W(J,j)}{\partial J}~~.\ee
The Legendre transformation of $W(J,j)$ with respect to
$J$ defines the $\sigma$-dependent effective potential \cite{5}
\be\label{2.13}
V(\sigma,j,J_0)=W(J(\sigma,j),j)+\frac{\sigma}{4}(J(\sigma,j)-J_0) ~~.\ee
It has an extremum at $\bar\sigma$, for which $J(\bar\sigma,j)=J_0$
where $J_0$ is the mass parameter in the Higgs potential.

Calculating $W(J,j)$ by the saddle point method
up to one-loop order, one gets
\ba\label{2.14}
&&W(J,j)=\frac{\lambda}{8}\hat\varphi^4_0(J,j)-\frac{J}{4}
\hat\varphi_0^2(J,j)-\frac{j}{2}\hat\varphi_0(J,j)\nonumber\\
&&+\frac{1}{(4\pi)^2}\frac{1}{2}\sum_pr_p\int^{\Lambda^2}
_0dK^2K^2\ln\left(1+\frac{m^2_p(J,j)}{K^2}
\right) ~~.\ea
The sum is over the particles of the standard model with their
$J$- and $j$-dependent masses. $r_p$ is a statistical factor
(3 for the $Z$, 6 for the $W$, -12 for the top, 1 for the Higgs).
For $j=0$ the result is gauge-invariant. To consider the
dependence of $W(J,j=0)$ on $J$ rather than on $j$ has the
additional advantage that for $J>0$ the Goldstone particles remain
massless and thus do not contribute. Furthermore, to one-loop
order, the potential $V(\sigma,J_0)$ remains real for all values
of $\sigma$. The derivative of $W(J)$ with respect to $J$
according to (\ref{2.12})
reproduces eq. (\ref{2.8}). As long as $\Delta\sigma$ is not very
small and $\Lambda$ is of order TeV or larger, $V(\sigma,J_0)$ 
calculated from (\ref{2.12}--\ref{2.14}), changes its shape 
significantly with a small change of $J_0$ for
$J_0$ near zero (and small $j$).

For the purpose of renormalization we can add to $W(J,j)$
a polynomial in $J$  up to second order. The linear piece
could be used to cancel the quadratic divergence in (\ref{2.8}).
However, it would then reappear in (\ref{2.9}). Instead, I will
normalize $\sigma$ such that it is zero in the limit of all 
particle masses going to zero, starting from $J<0$:
$\sigma(J\to 0_-,j\to0)=0$. To achieve this, we have to
subtract the $2\Lambda^2/8\pi^2$ part in (\ref{2.8}) and
(\ref{2.9}) by replacing $W(J,j)$ in (\ref{2.14}) by
$W(J,j)+\frac{\Lambda^2}{32\pi^2}2J$. A further change of
$W(J,j)$ using a subtraction term proportional to $J^2$ can 
remove the logarithmic 
divergence in (\ref{2.8}) and (\ref{2.13}). This subtraction can 
be interpreted 
as a renormalization of the Higgs coupling constant $\lambda$. 
Again, I do not perform such a complete subtraction since the 
corresponding term would then appear in (\ref{2.9}) where it has 
no physical basis. But we can remove the logarithmic divergence 
for regions of $J$ where the gauge bosons and the fermions remain 
massless. Since I am not completely certain about the necessity 
of this subtraction I will consider two cases: i) no subtraction 
proportional to $J^2$ and 
-- the more appealing one -- ii) a subtraction such that,  
besides the quadratic divergence, also the logarithmic divergence 
is removed in (\ref{2.9}). Accordingly, the factor which governs 
the logarithmic 
divergence of $\sigma$ and $V(\sigma,J_0)$ in the region of 
spontaneous symmetry breaking is ( to one loop order, and 
expressed in terms of masses)
\ba\label{2.15}
S_{Log\ \Lambda} = (\zeta m^4_H+3 m^4_Z+6m^4_W-12m^4_t)/
<\varphi_0>^4 ~~.\ea
$\zeta=1$ corresponds to no subtraction, while $\zeta = 0 $ is 
valid when the subtraction is performed according to ii)~.  
(I do not consider here the conventional non gauge invariant 
potential $V(<\varphi_0>,J_0)$. It would lead to $\zeta=3/2$).

As a speculation I will now assume a minimum influence
of new physics on the standard model. The idea is that the 
coupling constants of the standard model may take prefered 
values which stabilize the vacuum expectation value of the 
Higgs field and therefore also the particle masses.
In particular, $\Delta\sigma$ of eq. (\ref{2.10}) should
be independent of $\Lambda^2$, i.e. the
square bracket in (\ref{2.10}) should be proportional to
$1/\Lambda^2$, or vanish. If this is the case, the particle
couplings are not independent of each other, but satisfy
-- at least approximately -- the Veltman condition. 
Here one encounters the problem of the scale $(\mu)$ at which the
particle couplings should be taken. In particular, the Yukawa 
coupling of the top quark is sensitive to it. The vacuum expectation
value of the unrenormalized Higgs field is scale-invariant. But to
take advantage of this fact, higher order calculations, and a
knowledge of the scale dependence of $\Lambda$ ($\Lambda$ may
be related to high mass states) would be needed. However, it is 
clear that the natural scale for the couplings occuring in the 
loop integrals is $\mu \approx \Lambda$. 

Let us first 
assume that the cut off $\Lambda$ is very low and 
the cancellation of the coupling terms in 
(\ref{2.10}) occurs already at the weak scale of $\approx250~ GeV$, 
where the top mass is still big. 
Then, using (\ref{2.11}), and for the mass of the top $m_t(m_Z)=173$ GeV, 
the Higgs mass is predicted to
be $m_H(m_Z)\approx 280~GeV$.
Now we can 
take this value of the Higgs mass to look at the logarithmic 
divergence and calculate
$S_{Log\ \Lambda}$ .
Using $\zeta=1$ and again the scale $\mu \approx250~GeV$, we find
that $S_{Log\ \Lambda}=-0.2$~.
We thus have the surprising result that for a Higgs mass
of about 300 GeV the factors responsible for the quadratic
and the logarithmic divergence of the Higgs potential $V(\sigma,J_0)$
are both small. Let us now consider the extreme case
of strictly vanishing factors in front of $\Lambda^2$ and
$\log \Lambda$ for the one-loop Higgs potential $V(\sigma,J_0)$. 
This provides us with two equations which allow a calculation
of $m_t$ and $m_H$ in terms of the gauge couplings for $W$
and $Z$. The result is (for $\mu=250~GeV $ and a correspondingly 
very low cut off value)
\be\label{2.16}
m_t(m_Z)= 198~GeV~,\qquad m_H(m_Z)= 320~GeV~~.\ee
The fact that the value of $m_t$ obtained this way is not far away from
the experimental result is presumably a fortuitous coincidence, since for
the low cut-off considered a suppression of the logarithmic divergence 
does not appear plausible. 
But if not, it would indicate a very close connection of Higgs
and top with a Higgs mass not much different from $2m_t\approx
350$ GeV! On the other hand, the prefered value $\zeta = 0 $ gives here
no admissible solution.

Owing to the quadratic form of the two equations which simultaneously 
suppress quadratic and logarithmic divergences, another type of 
solution exists with smaller values for the Higgs mass. For this 
solution the relevant scale for the particles running in the 
loops must be extremly high 
(and hence the value of the cut off $\Lambda$) in order to get 
a large enough value 
for the top mass at the weak scale. We therefore take $\mu$ 
equal to the Planck mass, determine the Higgs and top couplings 
at this scale, and apply 
the two loop renormalization group equations to predict their values 
at the weak scale. This implies, of course, that physics beyond 
the standard model, which could influence the standard model 
couplings, can occur only near and above the Planck scale. 
The calculation, using $\alpha_s(m_Z)=0.12$ and $\zeta=1$, gives
\be\label{2.17}
          m_t(m_Z)=169~GeV~,~m_H(m_Z)=140~GeV  \ee
while for $\zeta=0$~,
\be\label{2.18}           
          m_t(m_Z)=168~GeV~,~m_H(m_Z)=137~GeV ~~. \ee
Both solution do not differ much since at the Planck scale $\lambda$
is found to be small (but not zero). I prefer the solution 
with $\zeta=0$. Firstly, it requires a very high scale which can be 
identified with a cut-off near the Planck mass. This cut-off is 
large enough that even the logarithm of it can provide for an 
order of magnitude suppression. Secondly, the divergencies are 
elminated in the unbroken phase as well and thirdly,  
$S_{Log\Lambda} $ can be expressed in terms of 
the $\beta$-function of $-\frac{J^2_0}{8\lambda}$,
i.e. the 
$\beta$-function of the zero order expression of $W(J_0)$ :
\begin{eqnarray} \label{2.19}
\displaystyle{4\pi^2~S_{Log\ \Lambda}} &=& 
\displaystyle{- \frac{\lambda^2(\mu)}{J^2_0(\mu)} 
\beta(\frac{J^2_0(\mu)}{\lambda(\mu)}) }\\[2mm] 
with ~~~\displaystyle{\beta(X)} &=& 
\displaystyle{\frac{\partial X}{\partial\ln\mu}~ .} \nonumber
\end{eqnarray}
In other words, the condition for the vanishing of the logarithmic 
divergence of $\bar \sigma$ can formally be viewed as a
requirement for a fixed point for $W_0(J_0)$.
 
The numbers obtained in 
(\ref{2.17}) and (\ref{2.18}) differ little from the result obtained 
by Bennett, Nielsen and Froggatt in the framework of their anti 
grand unification model. This model requires $\lambda(m_{Planck})=0$ 
(not far away from 
$\lambda(m_{Planck})=0.04$ obtained here), and the vanishing 
of $S_{Log\Lambda}$.

As a last point I like to comment on the question of the
possible participation of a 4th generation. If the particles of 
this generation
obtain their masses through the coupling to the standard model
Higgs particle, one cannot have $S_{\Lambda^2}$ 
and $S_{Log\ \Lambda}$
simultaneously sufficiently small or zero. The reason is
that fermion masses for the $b'$ and $t'$ would enter with masses
which are larger or roughly equal to the mass of the top.
One obtains imaginary solutions of the equations 
or, for $\zeta = 0$, too small values for the fermion masses.

We have seen that the possibility 
exists that the particles of the standard model have arranged their 
couplings such that the influence
from physics beyond the standard model is suppressed, stabilizing thereby
the particle masses. 
This is not a trivial statement. Furthermore, it can only occur for
the known three generations\footnote{Of course, nothing
can be said about particles with a different origin of their
masses.}. When we minimize simultaneously the quadratic and 
the logarithmic dependence of the vacuum expectation value of 
the Higgs field we find interesting 
relations between the Higgs and the top mass. For a very low, 
and perhaps too low, value of the cut-off scale we found 
$m_H \approx 2 m_t$. A much more interesting solution is the one 
with a very high value of the cut-off. In this case a properly 
subtracted form of $W(J)$ is taken such that there is no divergence 
in the parameter region of no spontaneous symmetry breaking. Then
the required absence of quadratic and logarithmic 
divergences in the physical region, and a cut-off value of the order 
of the Planck scale, leads to the correct value of the mass of 
the top and to a Higgs mass $m_H \approx 140~GeV$~. 

\section{Masses and Mixings of Quarks and Leptons}
\setcounter{equation}{0}

The possible intimate relation between Higgs and top
discussed in the previous paragraph also suggests a dominant
role of the top for the structure of the quark and lepton mass
matrices. The masses of the lighter particles can then be
expected to be related to the top mass by powers of a small
constant \cite{7,8}. Let us look at the quark and charged lepton masses
at the common scale $m_Z$ in the $\overline{MS}$ scheme (in GeV) \cite{9}
\be\label{3.1}
\begin{array}{lll}
m_t=173\pm6& m_b=2.84\pm0.10&
m_\tau=1.78\\
m_c=0.58\pm0.06& m_s=(70\pm14)10^{-3}&
m_\mu=106\times 10^{-3}\\
m_u=(2.0\pm0.5)10^{-3}& m_d=(3.6\pm0.8)10^{-3}&
m_e=0.51\times 10^{-3}\end{array}
\ee
We take as the small parameter $\epsilon\simeq\sqrt{\frac{m_c}{m_t}}
=0.058\pm0.004$. Then, the rule $m_t:m_c:m_u=1:\epsilon^2:\epsilon^4$
may be supposed to hold up to $O(\epsilon)$ corrections.
Similarly, the ratio of down quark masses and the ratio of charged lepton
masses are taken to be simple rational numbers times integer
powers of $\epsilon$. It is then plausible to use a corresponding
geometric structure also for the off-diagonal elements of the mass
matrices. Here, I will not go into details, since the paper
containing these suggestions is published \cite{8}. I will
only quote the results: The up-quark matrix can be taken real
and symmetric. For the down-quark and the charged lepton matrices
the simplest possible textures are used for which the 3rd
generation decouples from the first and second. The first and
second generation can mix by a complex entry. Taking this mixing
coefficient of the light generations purely imaginary, one obtains
-- for fixed $\epsilon$ -- a maximal CP-violation. It manifests
itself by making the unitarity triangle to be a right-angle
one: $\gamma\simeq90^o$. Besides the
top mass and $\epsilon$ there are only two additional parameters:
the beauty to top, and the $\tau$ to beauty mass ratios. With
these parameters one gets the following, so far quite successful,
numbers (in GeV):
\be\label{3.2}
\begin{array}{lll}
m_t=174& m_b=2.8&
m_\tau=1.77\\
m_c=0.58& m_s=84\times10^{-3}&
m_\mu=103\times 10^{-3}\\
m_u=1.95\times10^{-3}& m_d=4.2\times10^{-3}&
m_e=0.52\times 10^{-3}\end{array}
\ee
\be\label{3.3}
|V_{us}|=0.216,\ |V_{cb}|=0.041,\ |V_{ub}|=0.0034,\
|V_{td}|=0.0094,\ee
and
$\alpha=70^0,\ \beta=20^o,\ \gamma=90^o$.

\bigskip
It is a pleasure to thank Norma Mankoc Borstnik and Holger Nielsen
for the organization of the stimulating workshop in pleasant
surroundings. For lively and helpful discussions about the
possible Higgs-top connection I am much indebted to Ulrich
Ellwanger and Christof Wetterich.

\end{document}